\documentclass[aps,prl,preprint,groupedaddress,]{revtex4-1}

\usepackage{graphicx}
\usepackage{booktabs}

\begin{document}


\title{A Pulsed Muon Source Based on a High-Repetition-Rate Electron Accelerator}

\author{Meng~Lv}
\email{meng.lv@sjtu.edu.cn}
\author{Jiangtao~Wang}
\email{jiangtao\_wang@sjtu.edu.cn}
\author{Kim~Siang~Khaw}
\email{kimsiang84@sjtu.edu.cn}
\affiliation{Tsung-Dao Lee Institute and School of Physics and Astronomy, Shanghai Jiao Tong University, Shanghai, China}
\date{\today}

\begin{abstract}
Muons have established a unique and pivotal role in both fundamental physics and applied sciences. Given that a typical muon experiment spans roughly ten muon lifetimes, the optimal muon source should operate at around 50\,kHz in pulsed mode. However, existing muon facilities operate in either the 25-50\,Hz pulsed mode or continuous beam (DC) mode, which results in low-duty cycles for various muon experiments. As a result, precision muon physics with continuous muon beam has been limited by statistical uncertainty. In this study, we investigate the potential of a high-repetition-rate pulsed electron beam at the Shanghai SHINE facility to serve as a muon source driver. SHINE houses an 8-GeV CW superconducting RF linac, with a 1\,MHz bunch rate and 100\,pC bunch charge. Following X-ray production, the electron beam is deflected downstream of the undulators and absorbed in a beam dump. Using Geant4 Monte Carlo simulations, we estimated the yield of the muon beam to be approximately $10^{3}\mu^{\pm}$/bunch. This type of muon beam could be instrumental in a broad range of muon experiments, including muon lifetime measurement, a search for muonium to anti-muonium conversion, and the muon spin spectroscopy.
\end{abstract}


\maketitle

\section{Introduction}
Research in fundamental physics and applied science using muons has gained substantial interests in recent years. A long-standing discrepancy between theoretical predictions~\cite{Aoyama:2020ynm} and experimental measurements~\cite{Muong-2:2006rrc,Muong-2:2021ojo} regarding the muon's magnetic moment strongly suggests the presence of physics beyond the Standard Model of particle physics (for a recent review, see~\cite{Keshavarzi:2021eqa}). Techniques involving muon spectroscopy, such as muon spin rotation and muon-induced X-ray emission~\cite{Hillier:2022nat}, have catalyzed advancements in superconductivity, magnetism, and the elemental analysis of archaeological artifacts.

A common feature of existing and planned facilities is their reliance on high-power proton accelerators. Most of them operate in a multipurpose mode, where experiments with muons, neutrons, and pions are conducted simultaneously. These facilities currently operate either in a pulsed mode (25-50\,Hz, e.g., J-PARC in Japan~\cite{Miyake:2014zra}, and ISIS in the UK~\cite{Hillier:2018wkl}) or a continuous (DC) mode (e.g., PSI in Switzerland~\cite{Grillenberger:2021kyv}, TRIUMF in Canada~\cite{Marshall:1991wb}). This is also true for the five new muon facilities currently under study at CSNS~\cite{Vassilopoulos:2022ali}, HIAF/CiADS~\cite{Cai:2019feg}, RAON~\cite{Won:2014gja},  and FNAL~\cite{Gatto:2022olb}. Given that a typical muon experiment spans roughly ten muon lifetimes, current operating modes result in low-duty cycles for various muon experiments. For instance, precision muon physics with continuous muon beam has been limited by statistical uncertainty. Recently, several authors have noted that the optimal muon source for experiments such as the muon spin rotation (\textmu SR)~\cite{Cywinski:2009zz}, muon electric dipole moment~\cite{Adelmann:2010zz,Adelmann:2021udj}, and muonium to anti-muonium conversion~\cite{Willmann:2021boq,Kuno:2000wn} operates in pulsed mode with a repetition rate of several tens of kHz. A non-scaling fixed-field alternating gradient (FFAG) proton accelerator technology with a frequency of a few kHz~\cite{Kuno:2005zz} has been proposed for this purpose, but is still under development~\cite{Seidel:2021jyz}. The proton beam to be delivered to Fermilab's Mu2e experiment~\cite{Mu2e:2014fns} has a proton bunch repetition rate of 0.59\,MHz, achieved by resonantly extracting the proton bunch from a delivery ring. However, it is dedicated to the Mu2e experiment. Recent work at ORNL's SNS aimed to extract 50\,kHz proton pulses for \textmu SR applications, employing laser neutralization on a hydrogen ion beam~\cite{Liu:2020hcu}.

\begin{figure}[htbp]
    \centering
    \includegraphics[width=0.7\textwidth]{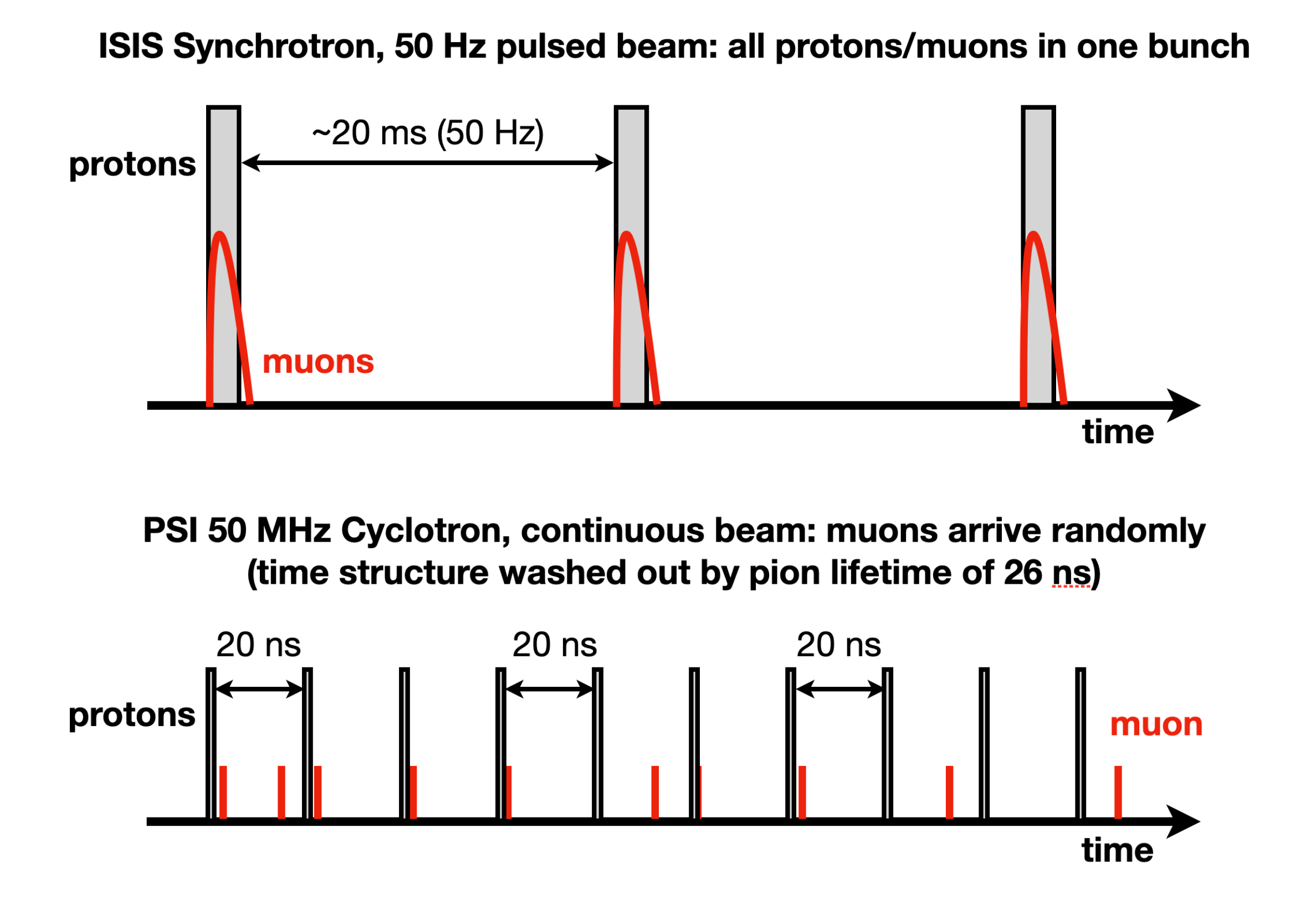}
    \caption{A comparison between pulsed and continuous muon beam.}
    \label{fig:MuonSourceTypes}
\end{figure}

Apart from the proton beam, an electron beam could also be used to drive a muon source. Recently, Nagamine \textit{et al.} proposed using a 300\,MeV, 10\,\textmu A electron accelerator for a \textmu SR facility~\cite{Nagamine:2009zz}, predicting a potential yield of $8 \times 10^{3}\mu^{+}$/s taking into account acceptance effects and transportation losses. A feasibility study for a muon beam using the Hall-A beam dump at the Continuous Electron Beam Accelerator Facility (CEBAF)~\cite{Dudek:2012vr} in Jefferson Lab has also been conducted~\cite{Fulci:2023hmx}. With a target electron number of about $10^{22}$/year, the anticipated muon flux is approximately $2\times10^{15}$/year. The latest advancements in laser wakefield acceleration (LWFA) technology have compactified electron accelerators to mere meters rather than kilometers. The development of such compact muon sources has been extensively studied recently~\cite{Titov:2009cr,Dreesen:2014mt,Rao:2018njj}, in anticipation of the imminent availability of high-repetition-rate femtosecond multi-PW lasers.

In this article, we explore the possibility of utilizing a high repetition rate electron beam at SHINE~\cite{Zhao:2018lcl} in Shanghai to power a muon source. We present a detailed analysis of the anticipated beam intensity for muons and other secondary particles for two distinct target configurations. Potential applications of these muon beams in the field of fundamental physics as well as applied science will also be discussed.

\section{MUON PRODUCTION BY ELECTRON}

Muon beams are typically produced as a tertiary beam within a proton accelerator complex. This production process is driven by a high-intensity proton beam hitting on a graphite target~\cite{Bungau:2014rxa,Berg:2015wna,Cook:2016sfz}. This results in the production of pions via strong nuclear interactions. Bending magnets are then used to extract these pions from the target area. As they travel through a lengthy decay channel, these pions gradually decay into muons, which are subsequently delivered to muon experiment zones. A particular category of muon, known as a surface muon, can be selected by ensuring the beam momentum is approximately 28 MeV/c. This specific muon beam originates from the pion decaying near the target's surface and is typically nearly 100\% polarized due to the parity-violating weak decay of the pion.

Contrarily, the muon production scheme for an electron-driver varies significantly. Muons can be created as a tertiary beam via photo-nuclear process. The photo-nuclear process entails the production of a real photon via the bremsstrahlung process, which is then followed by pion production through photo-excitation of the nucleus. Muons can also be created via the Bethe-Heitler process, which uniquely does not proceed through pion production and decay. The relative sizes of the cross sections are approximately in the ratio 1000:1 for photo-nuclear and pair production processes respectively, for electrons in the GeV scale~\cite{Blomqvist:1976mq,Nagamine:2009zz}.

\section{SHANGHAI SHINE FACILITY}

\begin{figure}[htbp]
    \centering
    \includegraphics[width=0.8\linewidth]{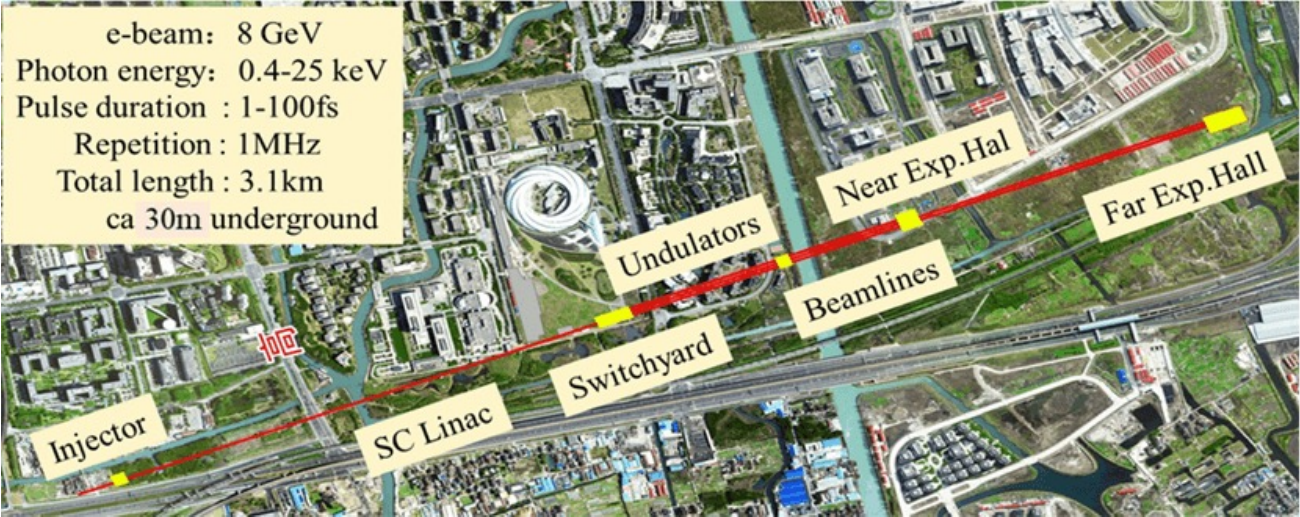}
    \caption{Aerial view of the SHINE facility~\cite{Zhao:2018lcl}.}
    \label{fig:SHINEFacility}
\end{figure}

The Shanghai High repetition rate XFEL and Extreme light facility (SHINE)~\cite{Zhao:2018lcl}, currently under construction at Shanghai's Zhangjiang High Technology Park (see Fig.~\ref{fig:SHINEFacility}), is a fourth-generation light source. This facility houses a 8-GeV CW superconducting RF linac, which delivers an 8-GeV bunched electron beam. The beam offers a repetition rate of up to 1\,MHz and a bunch charge of 100\,pC, culminating in an average current of 100\,\textmu A. SHINE is equipped with three undulator lines, capable of producing hard X-rays of up to 25\,keV. Following X-ray production, each electron beam is diverted and routed towards a beam dump~\cite{Xu:2020bd}. Interactions between the electron beam and either the beam dump or a thin target preceding the beam dump could generate muons and other secondary particles. 

\section{SIMULATION SETUP}

\begin{figure}[htbp]
    \centering
     \includegraphics[width=0.45\linewidth]{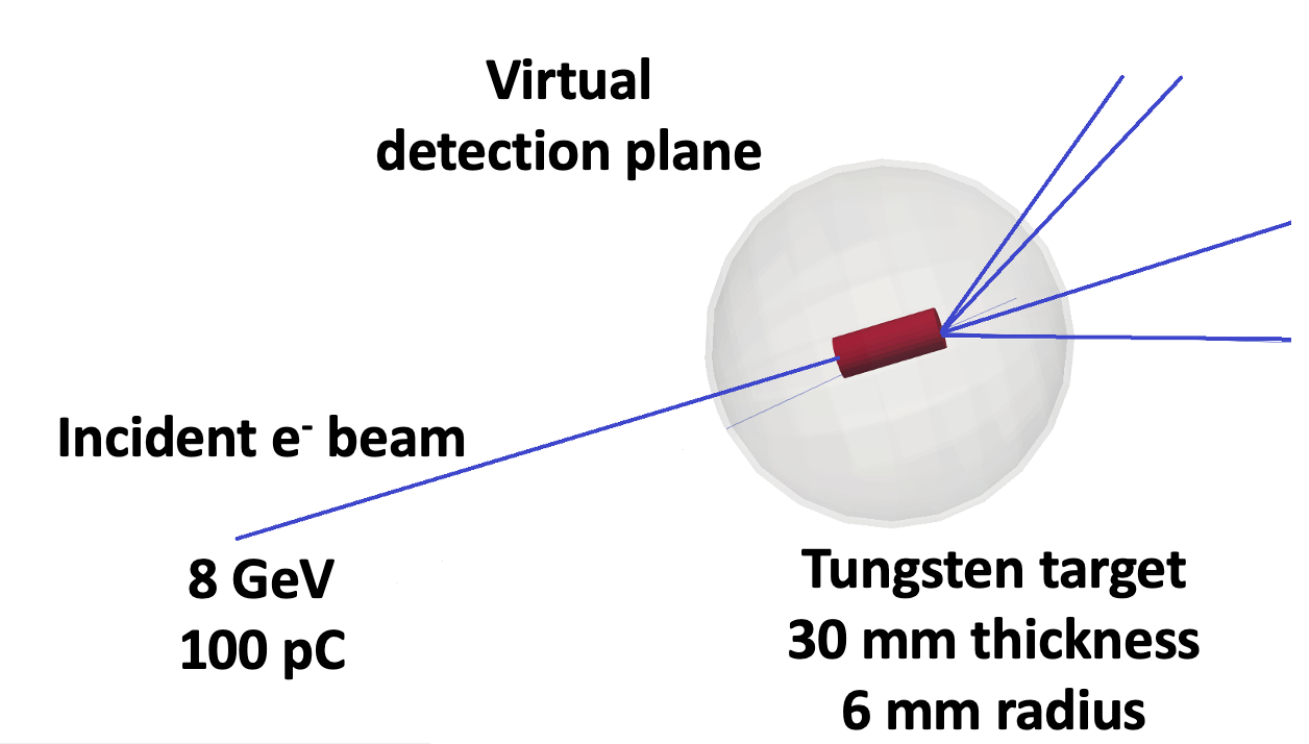}
    \caption{An event display from the Geant4 simulation for the target optimization study.}
    \label{fig:TargetOptimizationStudy}
\end{figure}

\begin{figure}[htbp]
    \centering
    \includegraphics[width=0.45\linewidth]{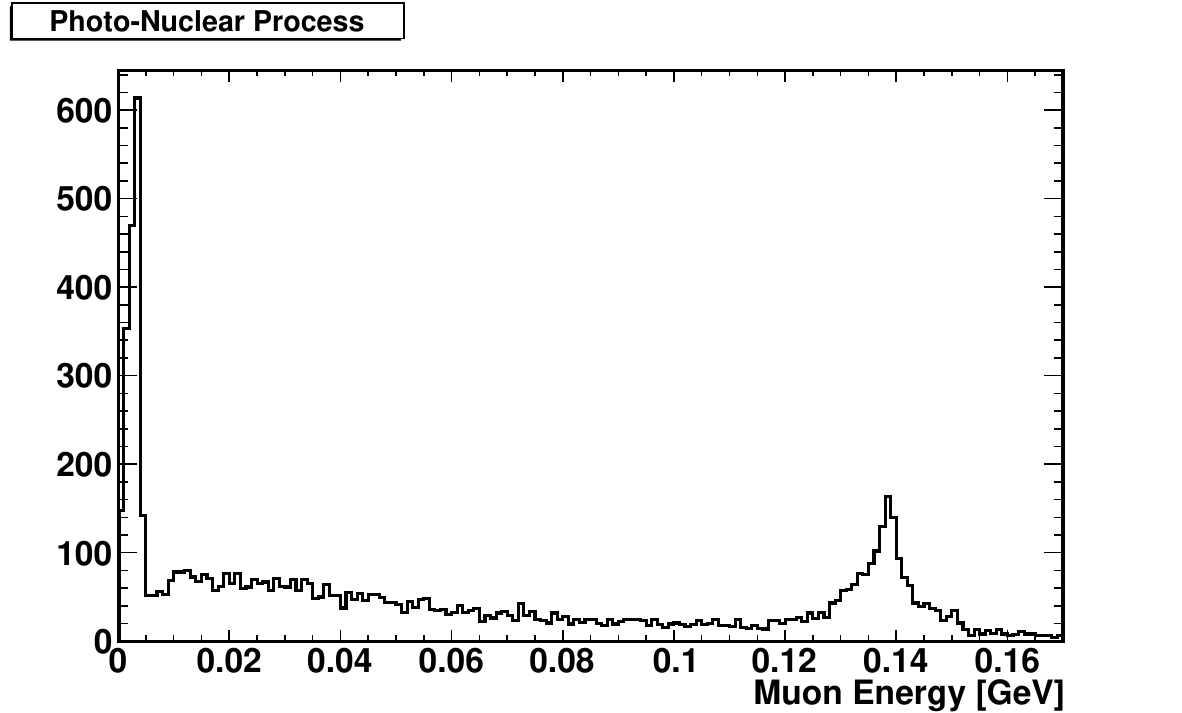}
    \includegraphics[width=0.45\linewidth]{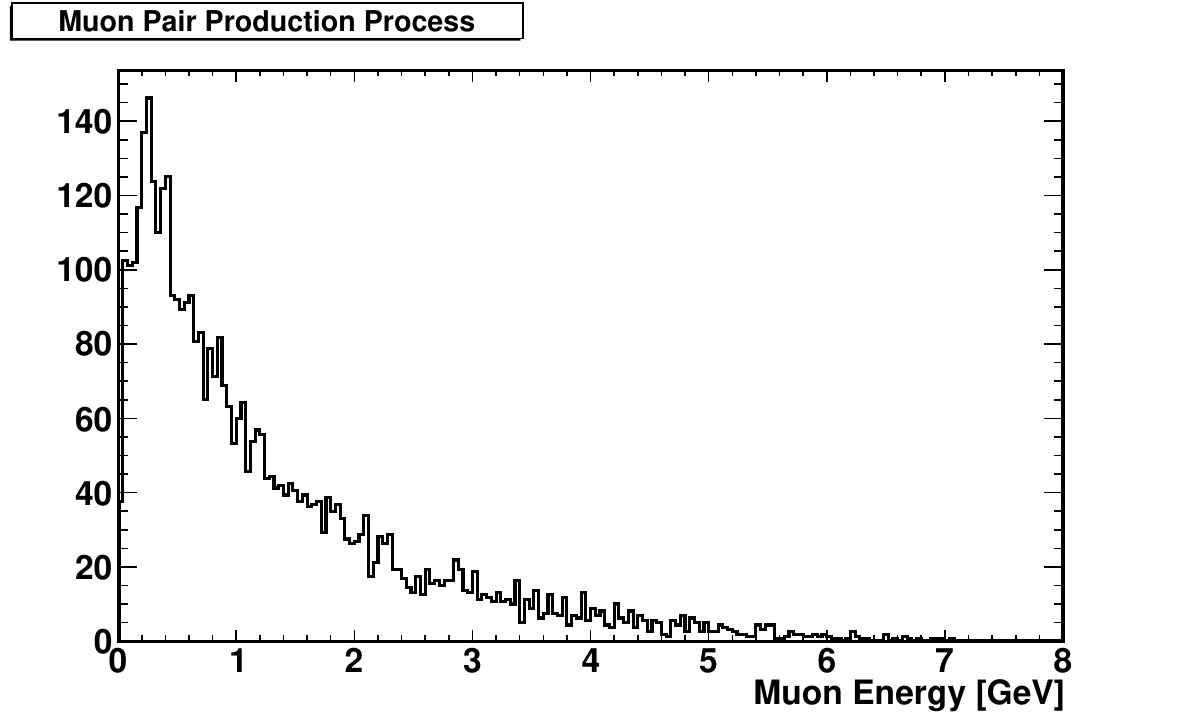}
    \includegraphics[width=0.45\linewidth]{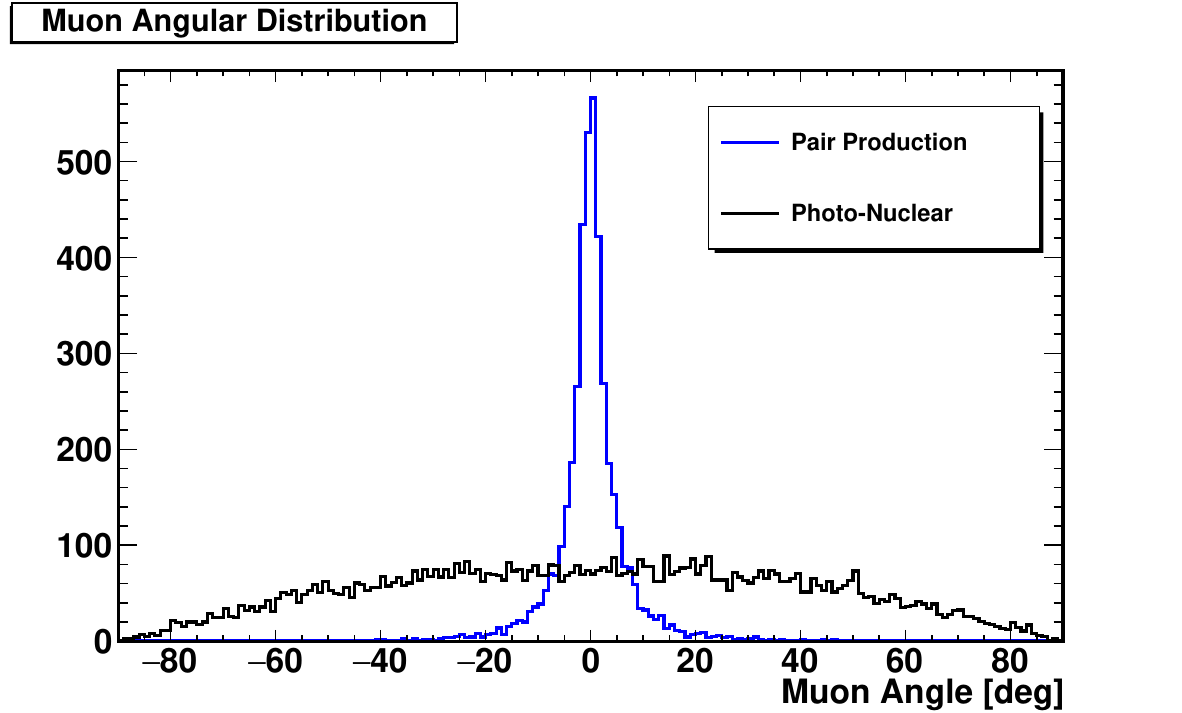}
    \caption{Distribution of muon energy and angle from both production channels per SHINE electron bunch of 100\,pC.}
    \label{fig:TargetOptimizationStudyResult}
\end{figure}

In order to simulate the production of muons, positrons, and pions resulting from electron interactions with the target, we employed the \texttt{musrSim} package~\cite{Sedlak:2012}, which is based on the Geant4 toolkit~\cite{GEANT4:2002zbu}. We modified the physics list to use the FTFP\_BERT model for hadronic processes. 
The initial phase of our investigation into the electron-beam-driven muon source focused on optimizing the muon production target. Based on prior study~\cite{Rao:2018njj}, we selected tungsten as our target material and designed it in a cylindrical shape. The target was optimized to maximize muon yield at a distance of 5 cm from the center of the target, as illustrated in Fig.~\ref{fig:TargetOptimizationStudy}. 
We found that the dimensions yielding the highest surface muon production for a tungsten target are a thickness of 30 mm and a radius of 6 mm. The energy and angular distribution of the muons are presented in Fig.~\ref{fig:TargetOptimizationStudyResult}. 
Muons generated from the photo-nuclear process exhibited a lower energy and a broader angular spread compared to those created from the pair-production process. This is consistent with our expectations since the former arise from pion decays. The yield from the photo-nuclear process is approximately $1.4 \times 10^{4}$ per bunch, while the yield from the pair production process is around $6.9 \times 10^{3}$ per bunch. Furthermore, we observed two distinct peaks in the momentum distribution, as depicted in Fig.~\ref{fig:TargetOptimizationStudyResult}(top). These peaks each correspond to the decays of pions and kaons, respectively, and have close to 100\% polarization, suitable for \textmu SR application. The yield of the surface muon at 4\,MeV is approximately 1,400 per bunch, resulting in an intense rate of $1.4 \times 10^{8}$/s for a 100\,kHz operation. Surface muons resulting from kaon decay present a unique prospect for probing extremely dense materials, owing to their higher penetration power~\cite{Grinenko:2023}.

\begin{figure}[htbp]
    \centering
    \includegraphics[width=0.7\linewidth]{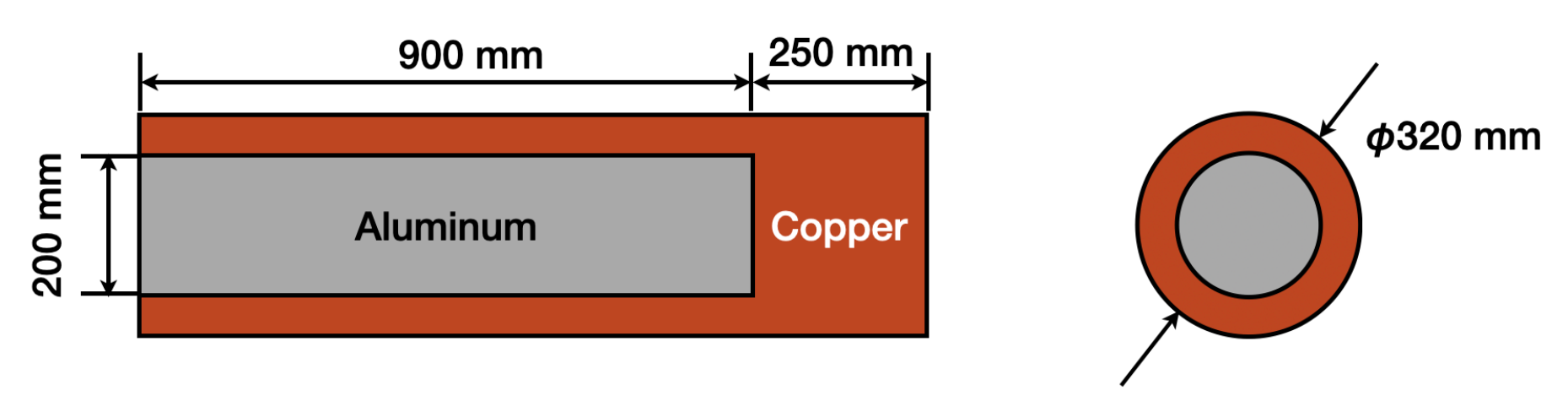}
    \caption{Geometry structure and materials of the SHINE beam dump. Adapted from~\cite{Xu:2020bd}.}
    \label{fig:shinebeamdump}
\end{figure}

In a scenario that minimizes modifications to the SHINE facility, we examined an alternative target configuration: the SHINE beam dump~\cite{Xu:2020bd}. As shown in Fig.~\ref{fig:shinebeamdump}, the SHINE beam dump is cylindrical and composed of aluminum and copper, with a radius of 16 cm and a total length of 115 cm. 

\begin{figure}[htbp]
    \centering
    \includegraphics[width=0.5\linewidth]{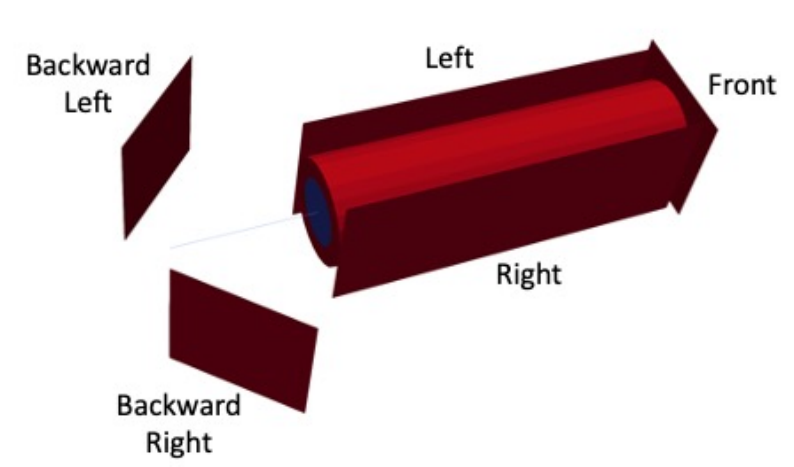}
    \caption{Placement of virtual detection planes for the SHINE beam dump configuration study.}
    \label{fig:EventDisplayBeamDump}
\end{figure}
Given the tiny emittance of the SHINE linac's electron beam, we approximated it as a pencil beam to simplify our model. To analyze the muon yield at various points around the beam dump, we installed five virtual detection planes: front, left side, right side, back left side, and back right side, as depicted in Fig.~\ref{fig:EventDisplayBeamDump}.

\begin{figure}[htbp]
    \centering
    \includegraphics[width=0.9\linewidth]{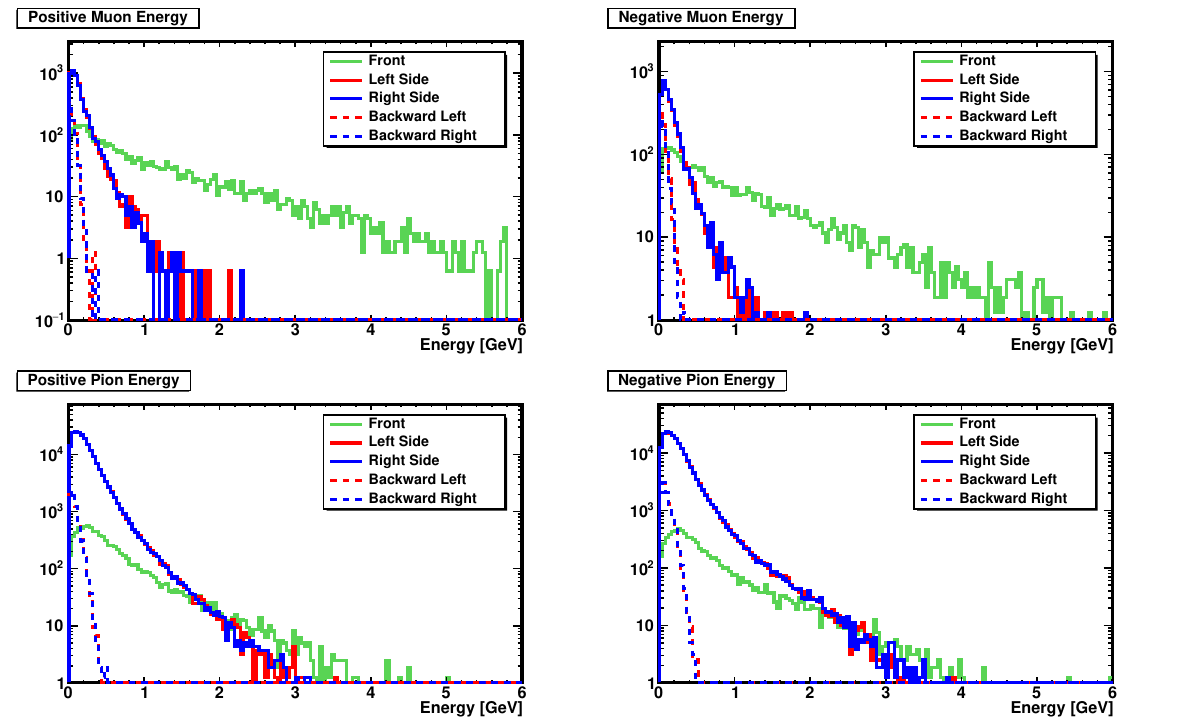}
    \caption{Energy distribution of the muons emitted from the beam dump, per bunch charge of 100\,pC at SHINE.}
    \label{fig:MuonAndPionRates}
\end{figure}

As anticipated, a substantial rate of muon and pion beams could still be detected from the beam dump. The energy distribution of the muons and pions is presented in Fig.~\ref{fig:MuonAndPionRates}, and the yield for each particle species is summarized in Tab~\ref{tab:muonpionyield}.

\begin{table}[htbp]
\centering
    \caption{Intensity of leptons and mesons produced from 8\,GeV electrons-on-target. The numbers here are normalized to the bunch charge of 100\,pC for SHINE.}
      \begin{tabular}{|c|c|c|c|c|c|}
        \hline
    Location & $\mu^{+}$/bunch & $\mu^{-}$/bunch   & $\pi^{+}$/bunch  & $\pi^{-}$/bunch \\
    \hline
     Front & $3.1\times10^{3}$ & $2.9\times10^{3}$ & $8.9\times10^{3}$ & $7.6\times10^{3}$ \\\hline
     Side (L) & $5.1\times10^{3}$ & $3.6\times10^{3}$ & $2.0\times10^{5}$ & $1.9\times10^{5}$ \\\hline
     Side (R) & $5.0\times10^{3}$ & $3.6\times10^{3}$ & $2.0\times10^{5}$ & $1.9\times10^{5}$ \\\hline
     Back (L)  & $6.6\times10^{2}$ & $7.5\times10^{2}$ & $6.8\times10^{3}$ & $1.0\times10^{4}$ \\\hline
     Back (R) & $6.5\times10^{2}$ & $7.1\times10^{2}$ & $6.6\times10^{3}$ & $1.0\times10^{4}$ \\
        \hline
    \end{tabular}
    \label{tab:muonpionyield}
\end{table}

\section{CONCLUSION}
In this study, we analyzed the feasibility of leveraging the high-repetition-rate electron beam at Shanghai's SHINE facility to generate a muon source. Through our simulations, we projected a muon yield of approximately $10^{4}\mu^{\pm}$/bunch using a cylindrical tungsten target with a thickness of 30 mm and a radius of 6 mm. In the beam dump configuration, we anticipate a yield of $10^{3}\mu^{\pm}$/bunch or even greater from all sides of the SHINE beam dump. Such a high-repetition-rate muon beam holds promise for diverse muon experiments, from muon lifetime measurements~\cite{Kanda:2022too} to the search for muonium-to-anti-muonium conversion~\cite{Bai:2022sxq}, as well as the application in muon spin spectroscopy studies~\cite{Li:2023gxn}. A dedicated study on the beam extraction will be performed as the next step of our research.

\section{ACKNOWLEDGEMENTS}
We would like to thank Guanghong Wang, Wenzhen Xu, Jianhui Chen, and Dong Wang from Shanghai Advanced Research Institute for useful discussions regarding the feasibility of developing a muon beam line using SHINE beam dumps. This work is supported by the Shanghai Pilot Program for Basic Research (21TQ1400221).


\end{document}